\RequirePackage{fix-cm}
\documentclass[smallextended]{svjour3}       % onecolumn (second format)
\smartqed  % flush right qed marks, e.g. at end of proof
\usepackage{graphicx}
 \journalname{ Experimental Astronomy }
\begin{document}

\title{Geant4 Simulations Of A Wide-Angle X-Ray Focusing Telescope}


\author{Donghua Zhao       \and
         Chen Zhang         \and
         Weimin Yuan         \and
         Shuangnan Zhang       \and
         Richard Willingale    \and
         Zhixing Ling          \and
}


\institute{Donghua Zhao (\email{zhaodh@bao.ac.cn}) \and Chen Zhang \and Weimin Yuan \and Zhixing Ling \at
           National Astronomical Observatories, Chinese Academy of Sciences, Beijing 100012, China;\\
           \and
            Shuangnan Zhang     \at
            National Astronomical Observatories, Chinese Academy of Sciences, Beijing 100012, China;\\
            Institute of High Energy Physics, Chinese Academy of Sciences, Beijing 100049, China;\\
            \and
            Richard Willingale  \at
			Department of Physics and Astronomy, University of Leicester, Leicester, LE1 7RH, UK. \\
}

\date{ Accepted:  DOI: 10.1007/s10686-017-9534-5}

\maketitle

\begin{abstract}
The rapid development of X-ray astronomy has been made possible by widely deploying X-ray focusing telescopes on board many X-ray satellites. Geant4 is a very powerful toolkit for Monte Carlo simulations and has remarkable abilities to model complex geometrical configurations. However, the library of physical processes available in Geant4 lacks a description of the reflection of X-ray photons at  a grazing incident angle which is the core physical process in the simulation of X-ray focusing telescopes. The scattering of low-energy charged particles from the mirror surfaces is another noteworthy process which is not yet incorporated into Geant4.

Here we describe a Monte Carlo model of a simplified wide-angle X-ray focusing telescope adopting lobster-eye optics and a silicon detector using the Geant4 toolkit.
With this model, we simulate the X-ray tracing, proton scattering and background detection.
We find that:
(1) the effective area obtained using Geant4 is in agreement with that obtained using Q software with an average difference of less than 3\%; 
(2) X-rays are the dominant background source below 10 keV; 
(3) the sensitivity of the telescope is better by at least one order of magnitude than that of a coded mask telescope with the same physical dimensions;
(4) the number of protons passing through the optics and reaching the detector by Firsov scattering is about 2.5 times that of multiple scattering for the lobster-eye telescope.

\end{abstract}

\section{Introduction}    
\label{sect:intro}

%As
With the rapid rise of space observational science in China, several space missions are in development. The Hard X-ray Modulation Telescope (HXMT), which will be launched in 2016-2017, is China's first dedicated astronomy satellite \cite{Zhang2009}. HXMT consists of three kinds of slat-collimated instruments and will perform a broad band (1-250 keV) X-ray all-sky imaging survey, as well as making pointed observations of X-ray sources to study their spectroscopic and temporal properties in detail. The missions of X-ray Timing and Polarization (XTP) and Einstein Probe (EP) \cite{Yuan2014,Zhao2014,Yuan2016} are another two X-ray astronomical satellites in development, and they will probably be launched around 2020. XTP will investigate matter under extreme conditions by accurate observation of light curves and energy spectra in energy band 1-100 keV, as well as measuring the X-ray polarization. The primary goals of EP are to discover X-ray transients and monitor variable objects in the energy band 0.5-4 keV with a very large instantaneous Field-of-View (FoV) of 60$\rm^o\times$60$\rm^o$. Unlike HXMT which employs only collimated instruments, XTP and EP will mainly adopt X-ray focusing telescopes.

The rapid development of X-ray astronomy has been made possible by widely applying X-ray focusing telescopes on satellites. 
The Wolter I \cite{Wolter1952} reflecting optics, consisting of a parabolic mirror followed by a hyperbolic mirror, is the most popular geometry employed. It has been applied aboard many satellites including HEAO-2 launched in 1978, Chandra and XMM- Newton in 1990s, Swift and Suzaku as well as NuSTAR in the 21st century. Although telescopes with Wolter I optics have remarkable spatial resolution (from sub-arcsecond to a few arcminutes), the FoV is usually very small (less than 1$\rm^{o}$). However, the lobster-eye optics \cite{Angel1979} overcomes the restricted FoV of Wolter optics. A telescope adopting lobster-eye optics can have a much wider FoV and lower mass at the same time, which make it very suitable as an X-ray all-sky monitor. Therefore, the lobster-eye optics has become a hot research topic and Monte Carlo simulations as described in this paper are an established and important method for the design and development of the optics.

Geant4 \cite{Agostinelli2003} is a very powerful toolkit for the simulation of the passage of particles through matter and has remarkable abilities to model complex geometrical configurations. It has been applied widely in high energy, nuclear and accelerator physics, as well as in space science. However, the library of physical processes available in Geant4 lacks the description of some physical processes at work, for example, the X-ray reflection at grazing angles and the scattering of electrons and protons on the polished mirror surfaces.

This paper is organized as follows: in Section \ref{sec:MCmodel} we will describe the Monte Carlo model of a wide-angle X-ray focusing telescope with specific parameters. In Section \ref{sec:Xray}, we will present the results of X-ray tracing with our model and the verifications of the results by comparing with the corresponding results with another model built with another tookit in detail. In Section \ref{sec:Proton} and \ref{sec:BKG}, we will briefly present the applications of Geant4 to the study of the scattering of low-energy protons at grazing incidence angles in the X-ray focusing telescope and that of the background, respectively.
Finally in Section \ref{sec:sum}, we conclude with a concise summary and a discussion of our investigation.

\section{Monte Carlo Model of A Wide-Angle X-Ray Focusing Telescope}
\label{sec:MCmodel}

Lobster-eye optics comprise very thin spherical plates in which there are numerous micro square pores, and the axes of the pores point radially to a common centre of curvature. The X-rays going through this type of optics will form a cruciform point-spread function \cite{Angel1979,Fraser1993} on the enclosed sphere with half the radius of curvature of the optics. 
Where the X-rays end up on the detector plane depends on how they pass through the micro pores. We suppose that the number of reflections off the two pairs of parallel walls are N$_x$ and N$_y$, respectively. If the micro pores are all perfectly oriented, the grazing incident X-rays undergoing only one reflection (or in general, N$_x$ and N$_y$, one is an even number and the other is an odd number) will be redirected to form two perpendicular foci, while those that undergo two reflections from adjacent walls (or in general, both N$_x$ and N$_y$ are odd numbers) will be brought to a focus, and the rays passing through the micro pores without reflections (or in general, both N$_x$ and N$_y$ are even numbers) will not be focused.
This kind of optics makes it possible to realize a wide-angle focusing telescope because of its imaging principle. 
As a result of the X-ray focusing, such a telescope will provide an order-of-magnitude or more improvement in sensitivity over current all-sky monitors. EP is just such a satellite which is designed to have a large instantaneous FoV of 60$\rm^o$ $\times$ 60$\rm^o$. 
The BepiColombo mission \cite{Johannes2010,Fraser2010}, which is due for launch in 2018, also incorporates square pore optics in its instrument Mercury Imaging X-ray Spectrometer (MIXS) but the lobster-eye optics on MIXS-C performs as a collimator with suppressed reflections within the pores.

\begin{figure}
\centering
\includegraphics[width=0.8\textwidth]{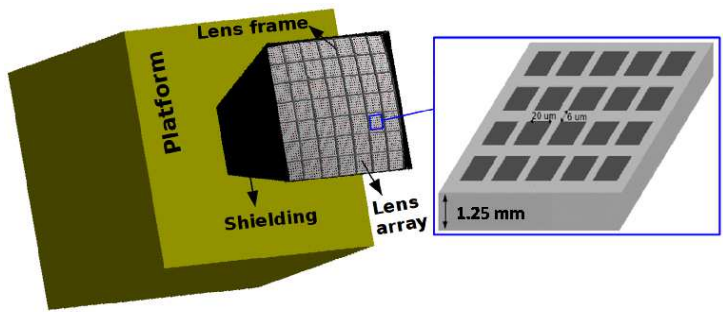}   %height=0.25\textheight,
\caption{(left) The mass models of an X-ray focusing telescope with lobster-eye optics (the size of micro pores is exaggerated for display).(right) A part of the mirror and its parameters. }
\label{Fig:MassModel}
\end{figure}

In this section, a Monte Carlo model of a wide-angle X-ray focusing telescope using lobster-eye optics is described.
The telescope in the simulations has a focal length of 375 mm. As shown in Figure \ref{Fig:MassModel}, this telescope consists of lobster-eye optics, a silicon pixel detector and some simple shielding as well as some supporting structures. The optics with an aperture of about 28$\times$28 cm$^2$ is composed of a 7$\times$7 spherical lens array and a supporting lens frame, which is similar to that of one module of the Wide-field X-ray Telescope (WXT) aboard the EP satellite \cite{Zhao2014}. Each lens has a thickness of 1.25 mm and a curvature of 750 mm. The micro pores in each lens have a size of 20$\times$20 $\rm \mu m^2$ and a length of 1.25 mm which is equal to the thickness of the lens.
The wall between two adjacent pores is 6 $\rm\mu m$ thick. These are the typical dimensions of a micro-square-pore lens. The material of the lens is lead glass comprising largely SiO2 and PbO. The walls of the pores are coated with 50 nm thick Iridium. 
The frame between two adjacent lenses has a thickness (or axial depth) of 10 mm and a width of 3 mm. Due to imperfection in the shaping, etching and coating processes, not all of the micro pores in the lenses point radially to the center of the curvature, and the coating surfaces on the walls of the micro pores have a certain microroughness. Both these factors are considered in the model. We set the volume of the silicon detector to 140 $\times$140 $\times$0.5 mm$^3$ and each pixel is 0.1$\times$0.1$\times$0.5 mm$^3$. The optics together with the detector produce a FoV of 20$\rm^o\times$20$\rm^o$. In the simulations, we take lead sheets with a thickness of 0.5 mm as shielding and supporting structures around the optics and the detector. We use a solid aluminium cube with a side length of 500 mm emulating the satellite platform.
The simplified shielding and platform, which can effectively reduce the photons and particles from the outside of the FoV, are just assumed in order to investigate the background in the telescope with Geant4.

We built the above simulation model based on the Geant4 toolkit (release 4.9.6p02).
The reflection of X-rays at grazing incidence angles, which is the core of the focusing process of X-rays, has not been included in Geant4.
To realize the X-ray focusing process, we utilize one extension of Geant4 named XRTG4, as introduced in Section \ref{sec:Xray}.
In order to simulate the scattering of particles at grazing incidence angles on the mirror surface, we use another extension which is presented in Section \ref{sec:Proton}.
With this Monte Carlo model, we examine the basic performance of the X-ray telescope, including realistic estimation of the collecting effective area as a function of X-ray energy, the point spread function and thus expected angular resolution, as well as the background noise on the detector. These studies are essential for better understanding of the instrument capability as well as for the shielding design.

%%%%%%%%%%%%%%%%%%%%%%%%%%%%%%%%%%%%%%%%%%%%%%%%%%%%%
%%%%%%%%%%%%%%%%%%%%%%%%%%%%%%%%%%%%%%%%%%%%%%%%%%%%%

\section{X-ray Tracing}
\label{sec:Xray}

Albeit including many physical processes, the Geant4 package lacks one for the X-rays at grazing incidence angles which is required in the simulations. Interestingly, Buis and Vacanti \cite{Buis2009} developed an extension (XRTG4) to Geant4 
allowing X-ray tracing simulations for optics of arbitrary complexity that can be described by the geometry library available in Geant4. 
The physical process with this extension is realized through the implementation of three classes: (1) G4XrayRefractionIndex manages the refraction index data for specific materials; (2) G4XraySurfaceProperty is a derived object of G4SurfaceProperty and it allows definition of the X-ray reflecting surface for grazing incidence X-ray scattering and to describe the microscopic surface details such as roughness; and (3) G4XrayGrazingAngleScattering implements the grazing angle scattering of X-rays on the surface during which the reflectivity changes with energy, incident angle and the surface roughness.

In our model, the surfaces between the vacuum in micro pores and the Iridium coatings on the walls of micro pores are defined as X-ray surfaces with G4XraySpecularReflectingSurface derived from G4XraySurfaceProperty. The roughness of the surface is specified. It is important to note that the refraction index of relevant materials must be given in the files named by material names in a folder called `dat' according to G4XrayRefractionIndex. G4XrayGrazingAngleScattering must be added to the process list of the particle of `gamma' in the corresponding physics list file.

The Q software\footnote{Q software is available from the author by request email zrw@le.ac.uk} is a sequential X-ray tracing package, which has been developed by R. Willingale at the University of Leicester, for simulations of different X-ray focusing telescopes including square Micro-Pore Optics (MPO). This package has been used for simulations of the conventional Wolter I shells like the Swift-XRT, Silicon Pore Optics (SPO) and the Wolter I optics implemented using square pore optics for MIXS-T aboard the Bepicolombo satellite \cite{Short2003,Spaan2008,Martindale2009,Willingale1998}. In this section, we make some comparisons with the relevant results obtained by using Q software to verify that Geant4 with XRTG4 can be applied to the simulations of X-ray reflection and scattering at grazing angles.
We also present some results of grazing angle reflection of X-rays utilizing Geant4.

\begin{figure}
\centering
\includegraphics[width=0.45\textwidth, height=0.25\textheight]{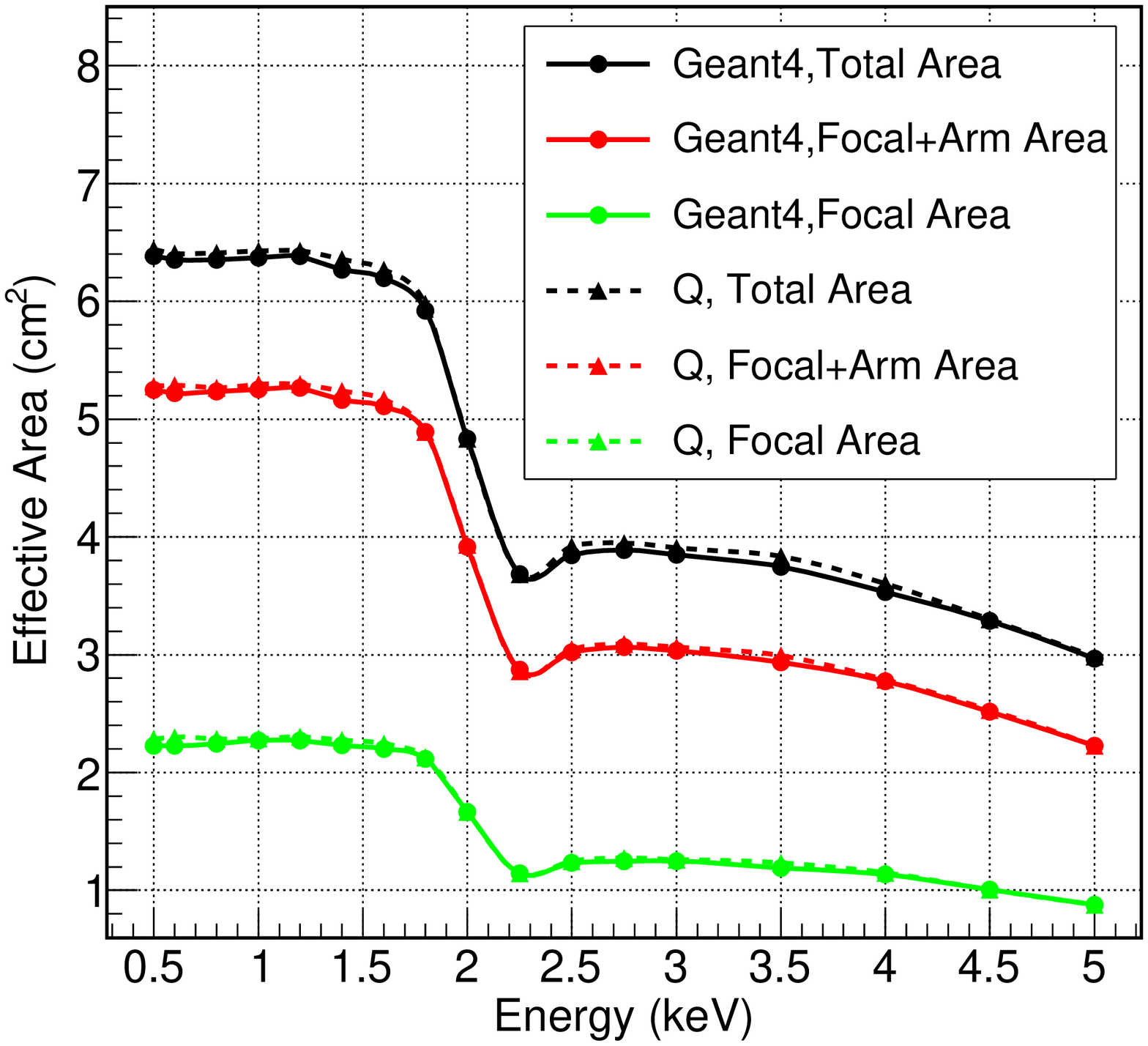}  
\includegraphics[width=0.45\textwidth, height=0.25\textheight]{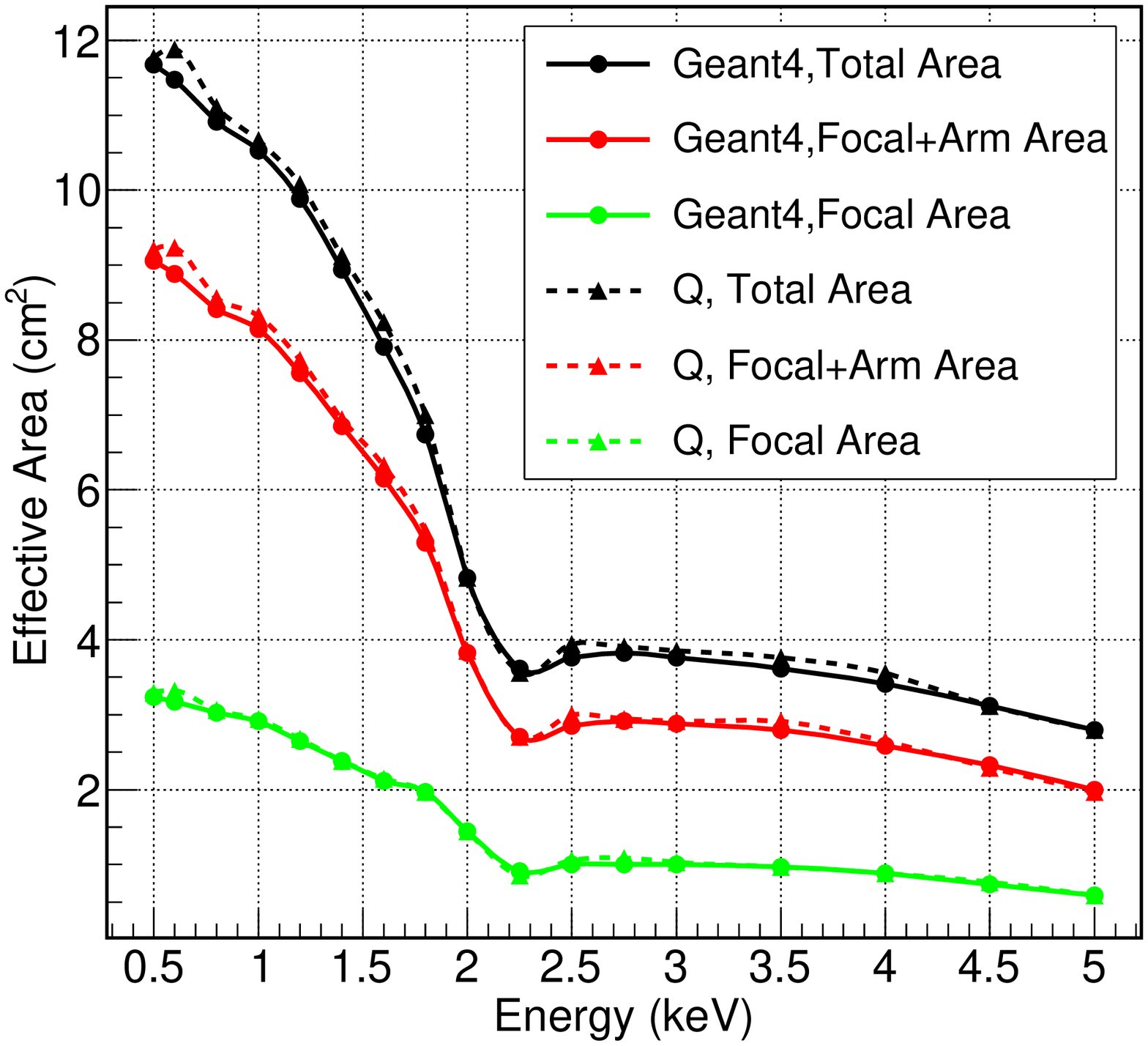}  
\includegraphics[width=0.91\textwidth, height=0.25\textheight]{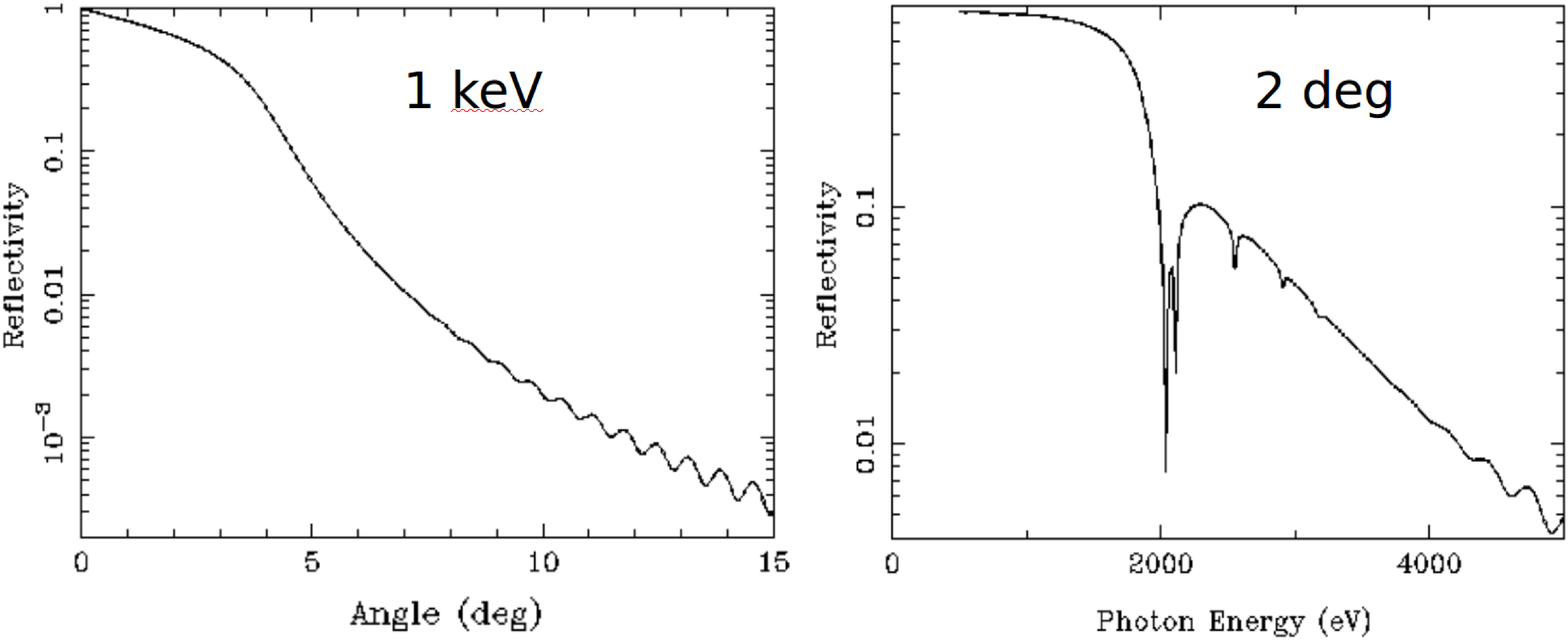}
\caption{Verification of Geant4 in simulations of the X-ray scattering at grazing incidence angles by comparing the effective area obtained from Geant4 (solid line with solid dots) and the Q software (dash line with solid triangles). (top left) the effective area varying with energy corresponding to one lens; (top right) the effective area varying with energy corresponding to the 7$\times$7 lens array; (bottom left) the reflectivity of X-rays at 1 keV on Ir coating surface changing with grazing incidence angles; (bottom right) the reflectivity of X-rays at a grazing incidence angle of 2$\rm^{o}$ on Ir coating surface changing with energy.}
\label{Fig:G4vsQ}
\end{figure}

The verification of the X-ray reflection simulations with Geant4 can be carried out by comparing the simulated effective area with that obtained with the Q software. For lobster-eye telescopes, the Point Spread Function (PSF) comprises three components arising from unreflected flux (diffuse transmission), single reflections and double reflections (or more generally an odd or even number of reflections). Therefore, we calculate three effective areas: the focal area from double reflections, the whole cross (focus and arms) area from both single and double reflections, and the total area for all of the three components. The comparisons of the results are shown in Figure \ref{Fig:G4vsQ} (top). 
For each energy point, 10$^7$ photons are simulated.
The top left panel of Figure \ref{Fig:G4vsQ} shows the results corresponding to the telescope with only one spherical lens in the center of its FoV. The average difference between the results of Geant4 and the Q software is less than 2.0$\rm\%$. When we use the whole lens array (7$\times$7 lenses) including supporting structure described in Section \ref{sec:MCmodel}, the average difference is less than 2.6$\%$ as shown in Figure \ref{Fig:G4vsQ} (top right). The effective area curve for 7$\times$7 lenses is much steeper than that for one lens below 2 keV. This is because the reflectivity decreases with the increasing incidence angle for a given energy, as shown in Figure \ref{Fig:G4vsQ} (bottom left), and decreases with the increasing photon energy for a given incidence angle, as shown in Figure \ref{Fig:G4vsQ} (bottom right).
The data in the two bottom panels are from the website\footnote{http://henke.lbl.gov/optical$\_$constants/layer2.html} and the relevant parameters include surface roughness=0, layer thickness=50 nm and the substrate material of SiO$_2$.
In the simulations, all the constants of refractive index used in Geant4 and Q software are from the same website\footnote{http://henke.lbl.gov/optical$\_$constants/getdb2.html}. 
The parameters of the optics are also the same in the two models. Both the surface roughness and the pointing deviation of the micro pores are zero. The only difference is that the supporting structure in Geant4 model has a thickness which can prevent some photons from going through the optics, while the supporting structure in Q software is just an abstract plane blocking photons. That is the main reason why the effective area obtained with Geant4 is generally smaller than that with Q software. 
We thus can conclude that the results given by the two software agree well with each other with an averaged deviation $<3$\% without considering the surface roughness. These comparisons verify the applicability of Geant4 and XRTG4 to the X-ray tracing simulations for lobster-eye telescopes.

Surface roughness can reduce the specular reflectance and cause diffuse reflection or scattering. Maximum peak-to-valley amplitude, surface spatial frequency, average roughness and root-mean-square (RMS) roughness are some frequently used parameters to describe it.
The different definitions of roughness is the main reason why we have not considered it when we compared the results between Geant4 and Q software. The X-ray surface in XRTG4 is characterized with only one parameter RMS, and the reflectivity is decreased by the factor
$\rm exp(-(\frac{4\pi\sigma sin(\theta)}{\lambda})^2)$,
where $\sigma$ is RMS, $\lambda$ is the wavelength, and $\theta$ is the grazing incidence angle.
At present, there is no module to describe the X-ray diffuse scattering from the rough surface in XRTG4 since this action is complicated. Therefore, some X-rays undergoing diffuse scattering may be lost.

\begin{figure}
\centering
\includegraphics[width=0.45\textwidth]{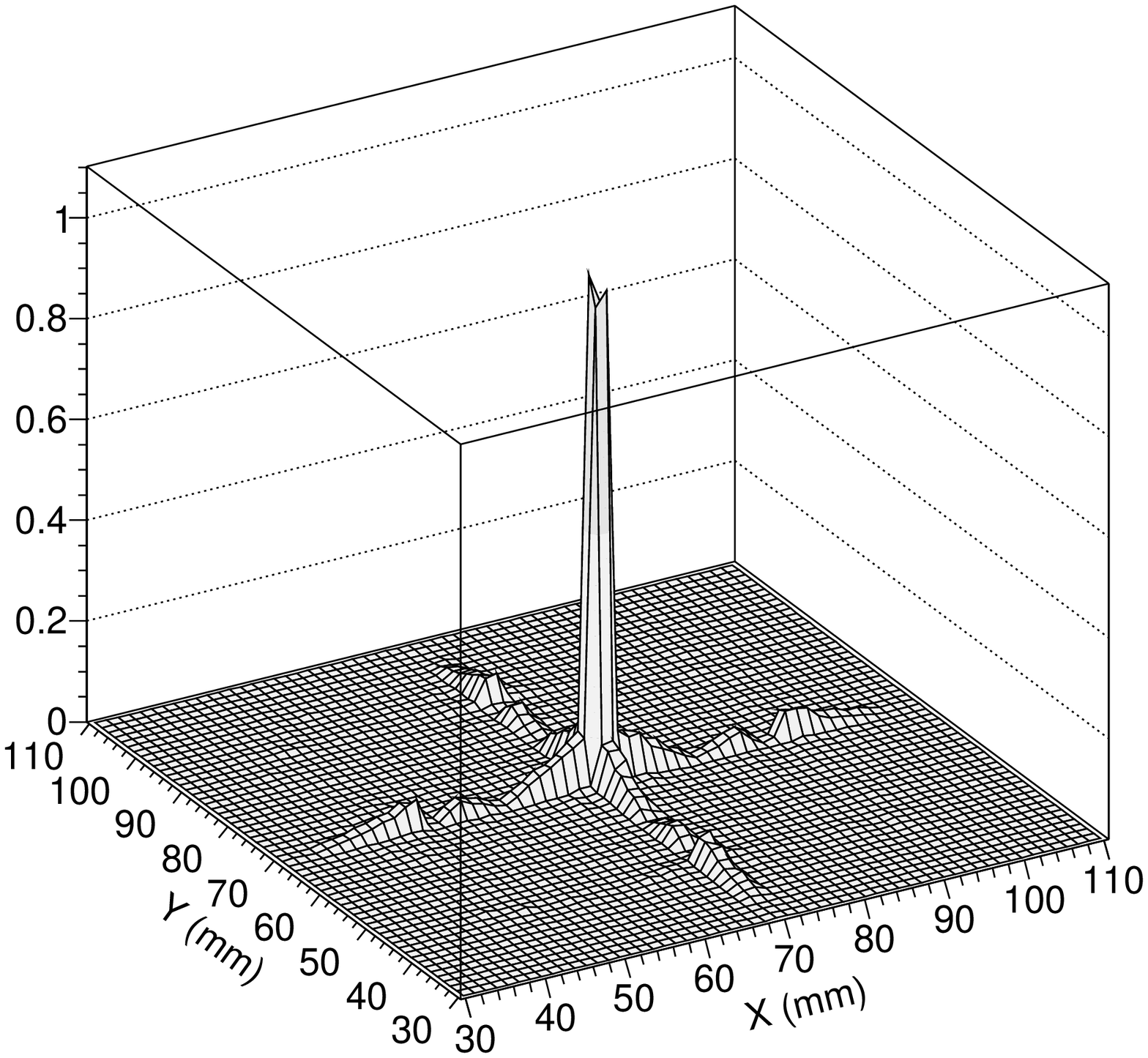}\includegraphics[width=0.45\textwidth]{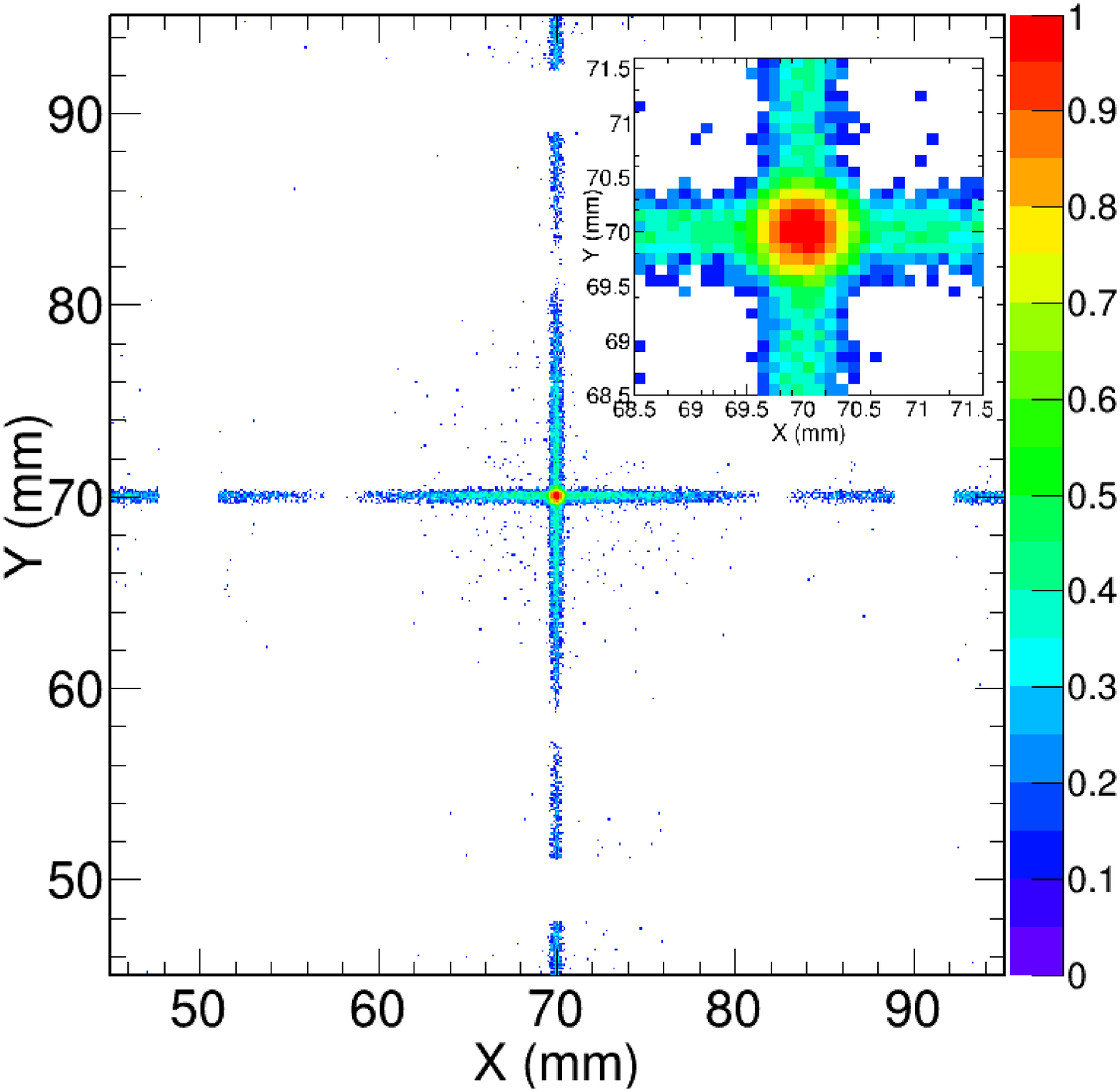}
\includegraphics[width=0.45\textwidth]{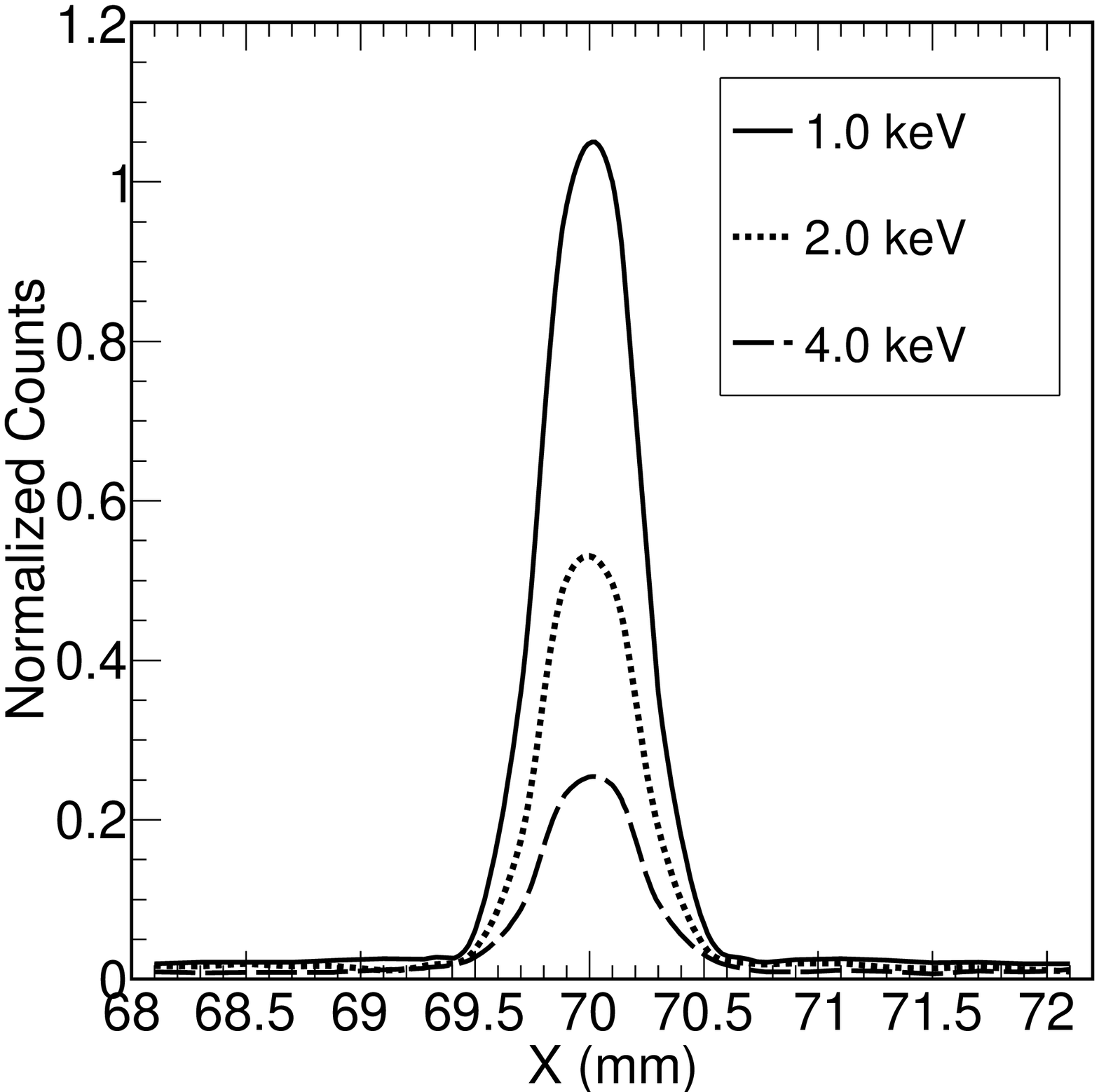}\includegraphics[width=0.45\textwidth]{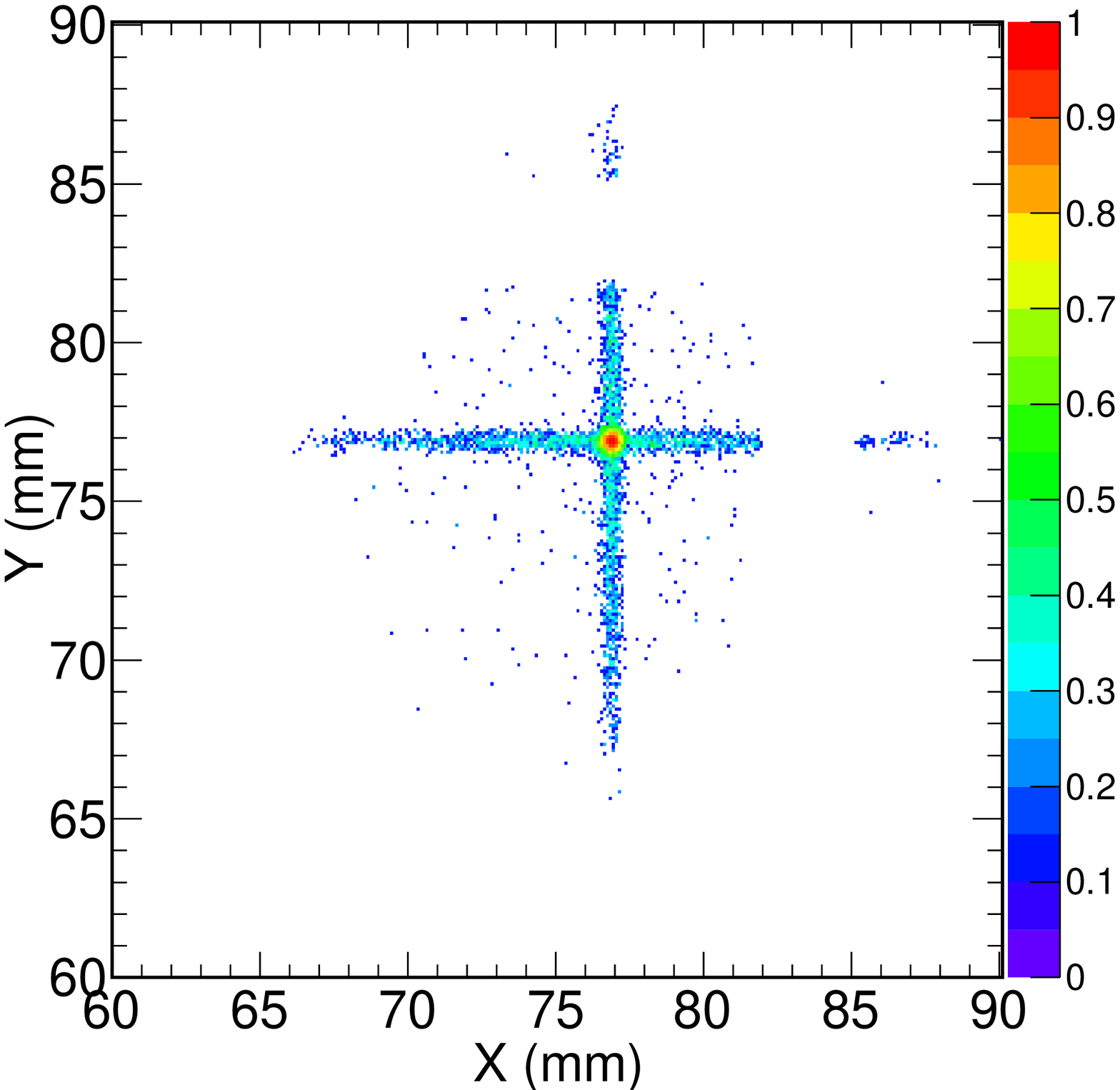}
\caption{The normalized PSF of the telescope described in Section \ref{sec:MCmodel} considering the surface roughness of about 0.55 nm and pore pointing deviation which follows a Gaussian distribution with $\sigma$=0.85 arcmin. Each PSF image is obtained by using 5$\times 10^{6}$ incident photons.
(top left) The surface plot of PSF at 1.0 keV. (top right) The contour plot of the PSF when the 1.0 keV parallel X-ray beam is on axis.
(bottom right) The PSF at 4.0 keV when the source moves to a position of 1.5$\rm^{o}$ zenith angle and 45$\rm^{o}$ azimuth angle.
(bottom left) One dimensional PSF at different energy considering one central group of data along the X-axis of the detector.
The color codes in the right two panels are in logarithmic scale.}
\label{Fig:PSF}
\end{figure}

It is difficult to avoid the surface roughness in the MPO production processes. Currently, the surface roughness can be reduced to less than 1 nm. After verifications of our program, we carry out more realistic simulations considering the surface roughness (RMS=0.55 nm).
We get the characteristic PSF of the lobster-eye telescope as shown in Figure \ref{Fig:PSF}. 
The size of the PSF structure decreases with the increasing photon energy.
We can learn from the left two panels in Figure \ref{Fig:PSF} that the photon count on the central focal area is much higher than that on other places and the photon count decreases with the increasing energy.
There are blank areas in the PSF (see the top right panel in Figure \ref{Fig:PSF}) around 50 mm and 90 mm, which is due to the shadow of the lens frame. The PSF and the shadow of the frame will move as the position of the source moves across the FoV.
For instance, when the source moves from the center of FoV (zenith angle,azimuth angle)=(0$\rm^{o}$, 0$\rm^o$) to a position with (zenith angle,azimuth angle)=(1.5$\rm^{o}$, 45$\rm^o$), the PSF and the shadow have a translation on the detector as shown in the bottom right panel of Figure \ref{Fig:PSF}.

In our simulations, a parameter which describes the pointing deviation of the long axes of micro pores (namely, the thickness dimension of lens) is also considered. It is assumed to follow a Gaussian distribution with a mean value of zero and $\sigma$ of 0.85 arcmin, which can lead to an angular resolution of about 4 arcmin. The detector plane is flat rather than spherical in our model. As a result, the angular resolution will be degraded at the edge of the FoV. In the method of simulations, we find that: the change in the angular resolution is less than 3\% in the central FoV of 5$\rm^{o}\times$5$\rm^{o}$; the angular resolution for a source at 10$\rm^{o}$ off axis is about 8.0 arcmin.

Based on these X-ray tracing simulations and the comparison with Q software as well as the verification completed by Buis and Vacanti \cite{Buis2009} for the conventional Wolter I and Silicon Pore Optics (SPO), we can conclude that the X-ray tracing simulations for grazing incidence X-ray focusing telescope can be successfully carried out with Geant4 and XRTG4.

%%%%%%%%%%%%%%%%%%%%%%%%%%%%%%%%%%%%%%%%%%%%%%%%%%%%%
%%%%%%%%%%%%%%%%%%%%%%%%%%%%%%%%%%%%%%%%%%%%%%%%%%%%%

\section{Proton Scattering}
\label{sec:Proton}

In addition to concentrating X-rays, a focusing telescope also acts as a concentrator of charged particles. 
These charged particles, which reach the detector by scattering through the optics, can degrade the spectral resolution and sensitivity, and even can shorten the detector lifetime. A significant increase of CTI (Charge Transfer Inefficiency) on ACIS (Advanced CCD Imaging Spectrometer) \cite{Prigozhin2000} aboard Chandra was observed soon after its launch in 1999. Some subsequent experiments found that this damage was due to the Non-Ionizing Energy Loss (NIEL) \cite{Dichter2003} caused by low energy protons (mainly in energy range 100-200 keV) at grazing incidence angles.
The passage of protons through the optics is an issue for focusing telescopes. In addition, electrons have similar scattering process \cite{Turner2006}. Therefore, it is very important to study the particle scattering to estimate the impact on detectors and to 
investigate what could be done to reduce the particles from the field of view. Both XMM-Newton and Swift-XRT \cite{XMM-Newton,Willingale2000,Burrows2005} included a magnetic diverter for deflecting charged particles (mainly electrons).

So far, the standard library of physics processes of Geant4 is not enough to describe the grazing scattering process of charged particles. Based on the theory of Firsov scattering \cite{Firsov1967}, Lei et al. (2004) \cite{Lei2004} developed an extension of Geant4 which can describe the scattering process of low-energy protons at grazing incidence angles. G4FirsovScattering is the core program of this extension. 
A much higher flux of protons (by a factor of 5 for XMM-Newton) will reach the detector using this extension than using Multiple SCattering process (MSC) of Geant4. The result with this extension is closer to the experimental result \cite{Lei2004}.
We modify and update this extension (FirsovSt) according to the Geant4 version used, and integrate it into our simulation model by adding this process into the physics list of proton and defining the surfaces between the Vacuum and the wall of micro pores as `firsov' surfaces. In the file of G4FirsovScattering, we can modify the available incidence angle and energy threshold as well as the distribution of energy loss.

\begin{figure}
\centering
\includegraphics[width=0.45\textwidth]{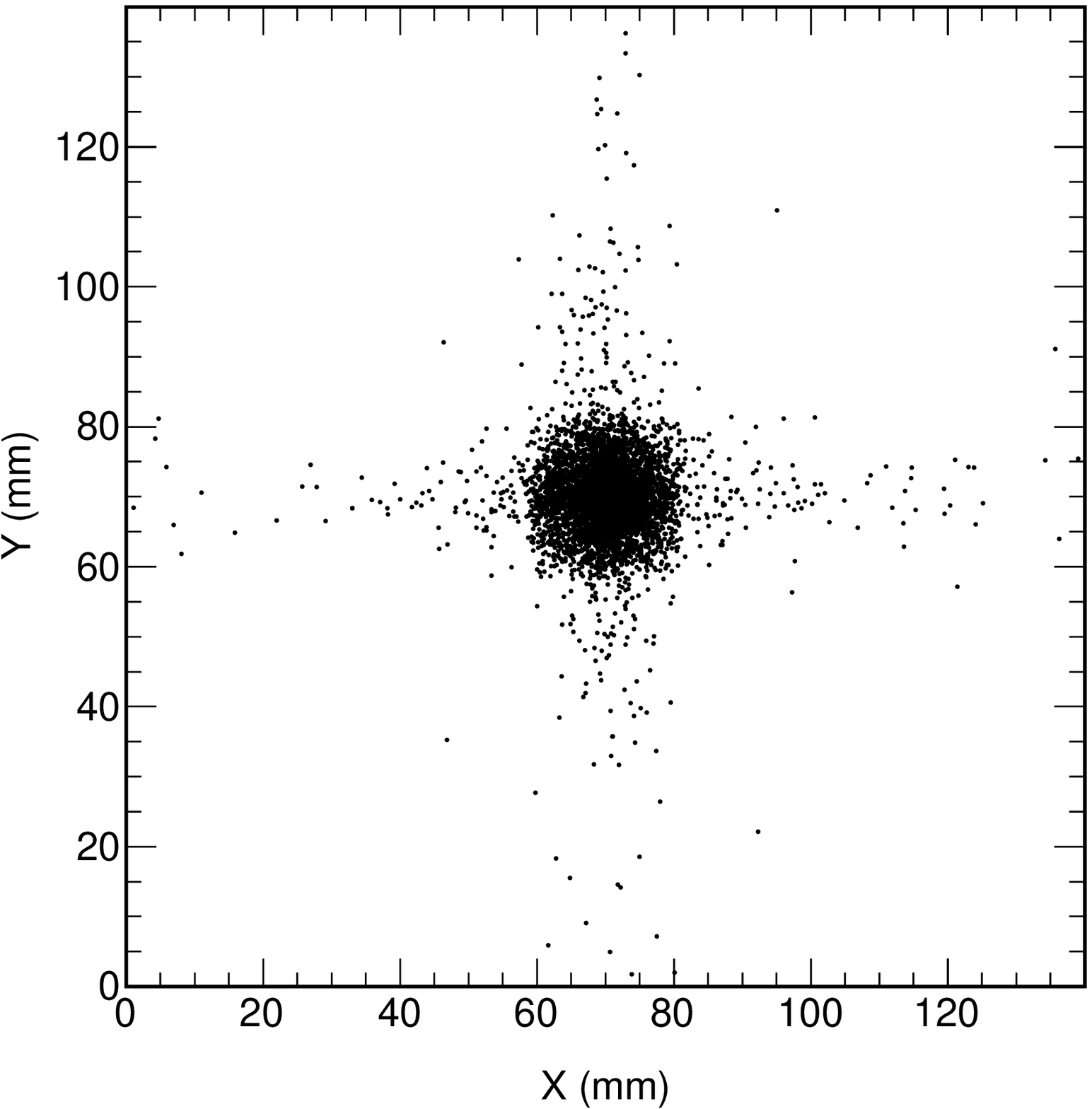}
\includegraphics[width=0.45\textwidth]{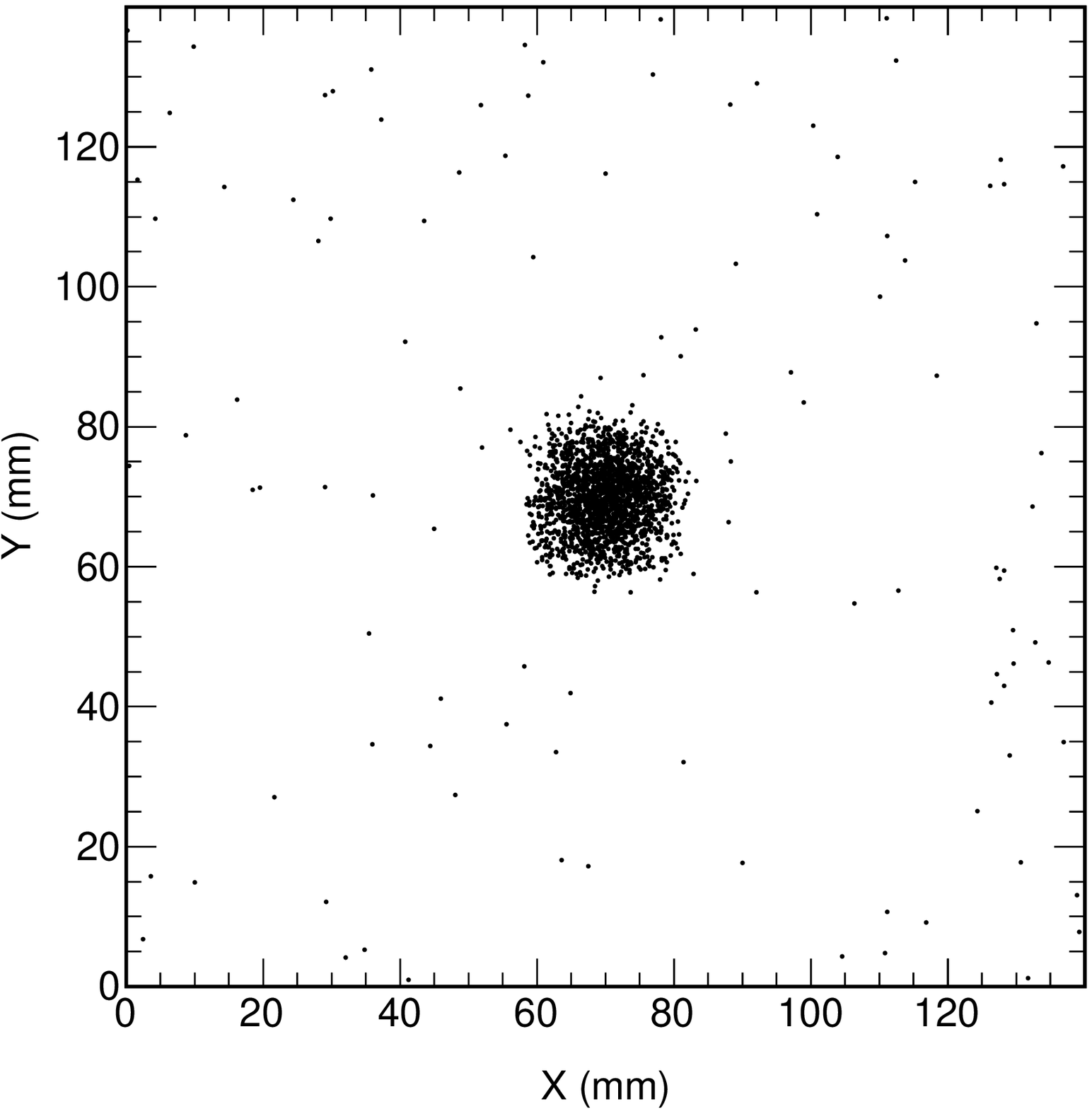}
\caption{The distributions of the protons, which are on-axis incident on the MPO optics and reach the detector, simulated with FirsovSt (left) and MSC (right). The number of protons in these simulations is 2.5$\times$10$^6$.}
\label{Fig:proton}
\end{figure}

For the telescope described in Section \ref{sec:MCmodel}, we simulate the scattering process of protons using FirsovSt and MSC, respectively. When we set parallel protons on axis, we get Figure \ref{Fig:proton}. We can learn from the figure that the behaviour of protons with FirsovSt is more like that of X-rays.
Base on this simulation, the protons reaching the detector with FirsovSt is about 2.5 times those with MSC.
This factor slightly decreases with the increase of focal length(150 mm: 2.7, 2m: 2.3) for the telescope which is similar to that described in Section \ref{sec:MCmodel}.

According to the experience of Chandra and XMM-Newton \cite{Nartallo2001}, different type of detector suffers effect of protons at different energy. For example, the FI (Front-Illuminated) MOS CCDs are mostly affected by protons at about 100 keV, the BI (Back-Illuminated) CCDs suffer most degradation from 1.3 MeV protons, and the thicker p-n CCDs are mostly affected by even higher energy protons. 
Estimating the count rate of the low-energy protons reaching the detector can provide data which impacts on the design of the detector.

For the telescope described in Section \ref{sec:MCmodel}, we do simulations with FirsovSt using the kappa energy spectrum \cite{Petrov2008} of proton flux near the geomagnetic equator in low earth orbit.
\begin{equation}
\label{eq:Kappa}
f(E) = A[1+\frac{E}{kE_0}]^{-k-1}	
\end{equation}
where $A$ = 328 cm$^{-2}$ s$^{-1}$ sr$^{-1}$ keV$^{-1}$, $k$ = 3.2 and $E_0$ = 22 keV.
These protons are produced by the charge-exchange mechanism and are
different from the trapped protons described in NASA radiation belt models. In simulations, the energy of protons is 30-10000 keV and the incidence direction is isotropic. We only take into account the protons which pass through the optics and finally reach the focal detector. The results are shown in Table \ref{Tab:ProtonRate}. The number of protons for each energy range is 2$\times$10$^8$ in the simulation. We also estimate the count rate with the process of MSC for the sake of comparison. The results in the table show that the count rate with FirsovSt is obviously higher than that with MSC. According to the data on the NIST website\footnote{http://physics.nist.gov/PhysRefData/Star/Text/PSTAR.html},
the projected ranges of protons in Silicon with 30 keV, 8000 keV and 10000 keV are 302 nm, 0.5 mm and 0.7 mm, respectively. Therefore, almost all the protons between 30-8000 keV will deposit energy in 0.5 mm thick silicon detector. The protons with energy above 8000 keV will deposit energy above 198 keV since the stopping power of proton in Silicon for minimum ionizing particles is about 1.7 MeV$\cdot$cm$^2$/g.
It should be noted that the whole silicon detector is the sensitive area in our model.
However, some structures such as a coating film will be placed in front of the sensitive area of a real detector to stop low energy photons. For example, Aluminium of 150 nm thickness is a typical coating film for a semiconductor detector.
Based on the data of NIST website, the protons below 13 keV can be obstructed by this film and the protons above 25 keV will pass through it with residual energy above 10 keV. Therefore, the protons between 13 and 25 keV will penitrate the film and eventually deposit energy below 10 keV in the silicon detector, which will cause additional background. 
Fortunately, the protons with energy below 30 keV can be deflected by a diverter which is often employed in X-ray focusing telescopes.
Therefore, the low-energy protons have negligible impact on the background below 10 keV for the telescope described in Section \ref{sec:MCmodel}.

\begin{table}
\begin{center}
\caption{The count rate of low-energy protons in different energy ranges based on simulations using the kappa energy spectrum (Equation (1)) for the telescope described in the Section \ref{sec:MCmodel}.}
\label{Tab:ProtonRate}
\small
\begin{tabular}{|c|c|c|c|c|c|}
\hline
Energy (keV)            & 30-100 & 100-500 & 500-1000 & 1000-10000 \\
\hline
Firsov(counts/s/cm$^2$) 	&2.270 	&0.551	&0.010	&0.002 \\
\hline
MSC(counts/s/cm$^2$)    	&0.984	&0.216	&0.004	&0.001 \\
\hline
\end{tabular}
\end{center}
\end{table}

The physical mechanism that makes optics behave as a particle concentrator is not yet well understood. The Firsov scattering and the multiple scattering are two of the theories which can be used to interpret the proton reflections off mirrors. In fact, these two theories may compete with each other, or their mixing ratio may be a function of the roughness of the surface. However, currently there is no information available about their mixing ratio and about the relationship between the mixing ratio and surface roughness. The model should be updated with the theory development and with new experimental data.

\section{Background and Sensitivity}
\label{sec:BKG}

Geant4-based models are in fact already widely used to study the effect of cosmic radiation on the spacecraft structures and instruments. For example, Geant4 has been used to estimate the background of many missions such as MAXI \cite{Hiroshi2003}, SVOM \cite{Zhao2012}, LOFT \cite{Campana2013}.

For the mass model described in Section \ref{sec:MCmodel}, we estimate its background level when it is in the Low Earth Orbit (LEO, an altitude of 600 km). In the simulations, we consider both electromagnetic interactions and the hadronic interactions. The standard electromagnetic interactions with the currently optimized models are used and the fluorescence X-rays, Auger electrons as well as particle induced X-ray emissions are taken into account.For the hadronic physics, different models are used in different energy ranges (for example, Bertini Cascade model, Quark-Gluon String model and so on). Both elastic and inelastic processes are considered.

For the input sources of background, we use the energy spectra of cosmic X-rays (thereafter CXB) and albedo $\gamma$-rays, which were adopted in simulations of SVOM \cite{Zhao2012}. We also consider the spectra of different particles, such as protons, positrons and electrons, etc, which were used for the background simulations of LOFT \cite{Campana2013}. 
SAA passages have not been considered for the moment. 
For soft X-ray telescopes, the soft X-rays from the Galaxy can not be neglected, which mainly include the thermal components probably from the Local Bubble and the Galactic halo. In simulations, we use the soft X-ray spectrum from the Lockman Hole field \cite{Miyaji1998}. %as the Galactic components. 
The X-rays below 1 keV are easily absorbed by the interstellar gas. We assume that the interstellar gas has a hydrogen column density of 3$\times$10$^{20}$ cm$^{-2}$. In the simulations, we consider the shielding effect of the Earth for the background and set the Earth behind our instrument. The calculation according to the radius of the Earth and the altitude shows that the Earth and the atmosphere are totally outside the FoV. Therefore, we do not take into account the fluorescent X-rays from the atmosphere since their energy is generally low and they hardly penetrate the shields we set up. As a result of the shielding structure, the background induced by albedo $\gamma$-rays from the atmosphere is very low.

\begin{figure}
\centering
\includegraphics[height=0.3\textheight, width=0.7\textwidth]{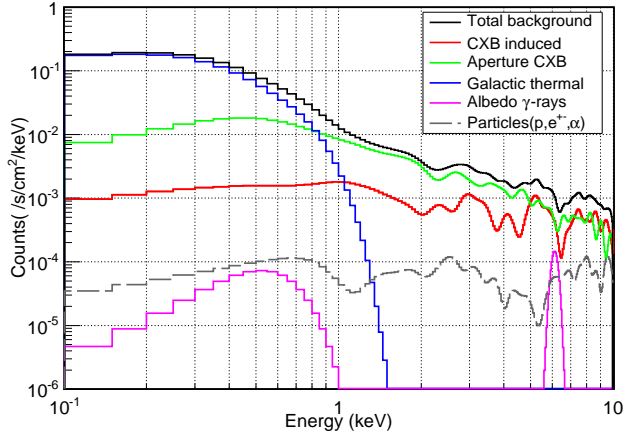}  
\caption{The background estimated with Geant4 for the instrument described in Section \ref{sec:MCmodel}.  }
\label{Fig:BKG}
\end{figure}

We finally obtain the background as shown in Figure \ref{Fig:BKG} for the telescope described in Section \ref{sec:MCmodel}. In the energy range of 0.1-1 keV, about 87\% of background is due to the thermal component of Galaxy. Between 1-10 keV, the background caused by the aperture CXB and non-aperture CXB are about 67\% and 28\%, respectively; the particles-induced background is less than 3\%. Therefore, we can conclude that X-rays play the dominant role in the background sources below 10 keV for such a soft X-ray focusing telescope.

As the first case study, we set a simple Pb sheet as the surrounding shield; the work to design a more realistic shielding configuration is in progress. For the current structure of telescope, the background count rate is 0.027 cts/s/cm$^{-2}$ in the energy range 0.5-4 keV. Based on such an estimation, the on-axis sensitivity is expected to be 0.02 mCrab (5$\sigma$ detection level, in 10 ks, for a source with a power law spectrum with photon index 2.1 and absorption 3$\times$10$^{20}$ cm$^{-2}$).

\begin{figure}
\centering
\includegraphics[width=0.4\textwidth,height=0.3\textheight]{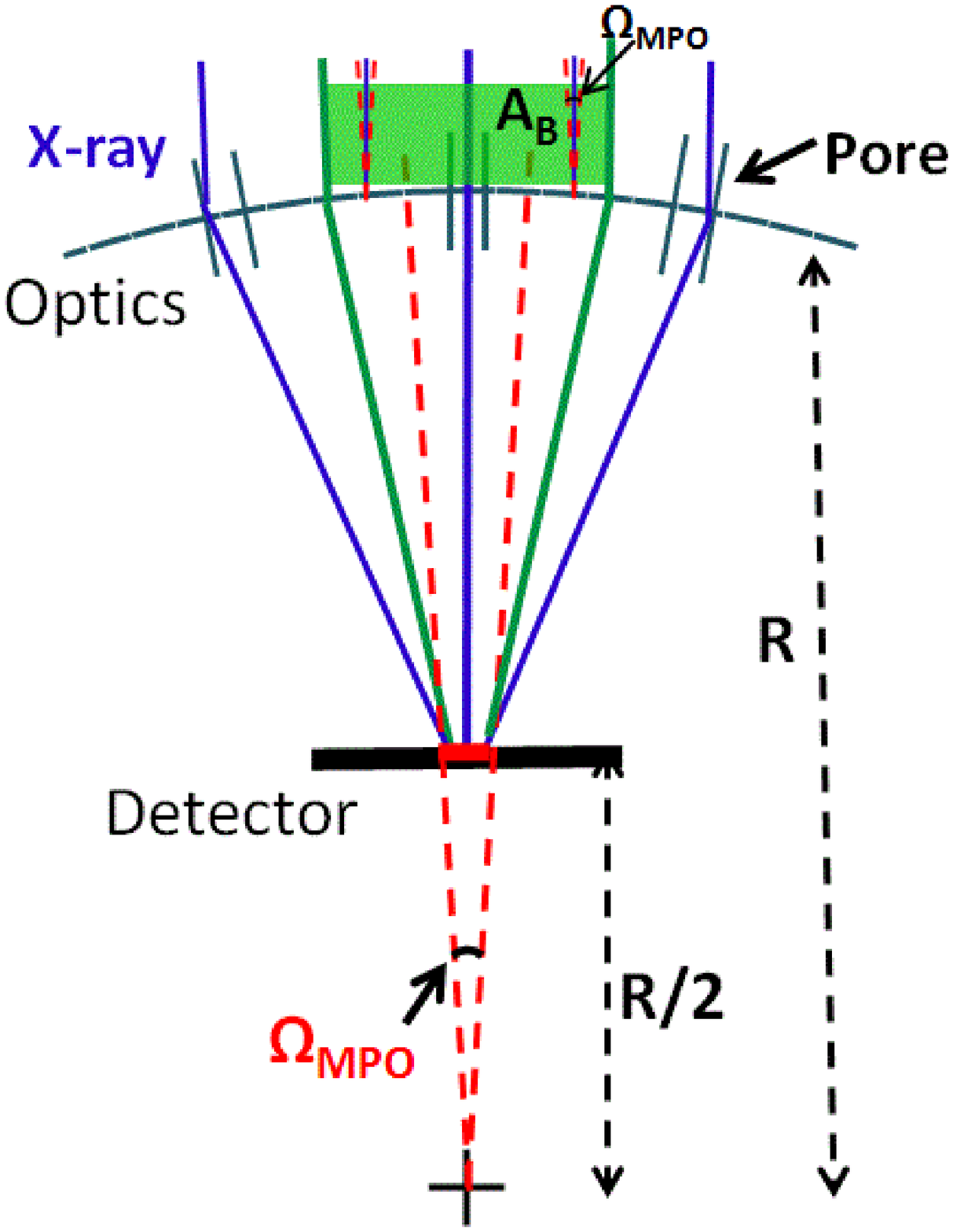}
\includegraphics[width=0.4\textwidth]{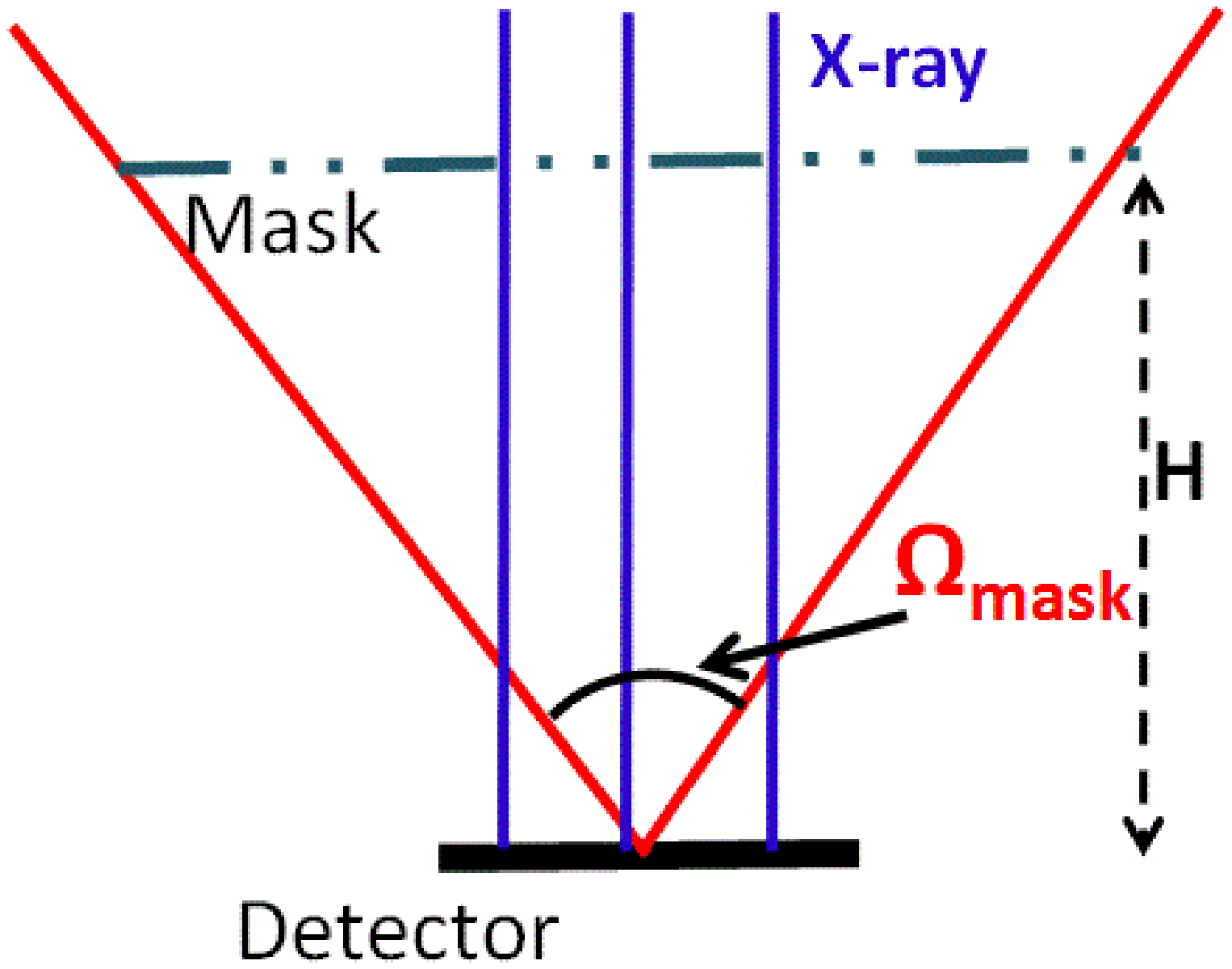}
\caption{Schematic views of the MPO focusing telescope (left) and the coded mask telescope (right). The scale is exaggerated.}
\label{Fig:MPOandMask}
\end{figure}

We suppose there was a coded mask telescope which had the same detector (14 cm$\times$14 cm) and the same aperture dimension (28 cm$\times$28 cm) as the MPO telescope described in Section \ref{sec:MCmodel}. As shown in Figure \ref{Fig:MPOandMask}, the distance H from the mask to the detector is equal to the focal length of MPO telescope R/2. Thus the coded mask telescope has a FoV of about 40$\rm^{o}\times$40$\rm^{o}$. The coded mask telescope discussed here is an instrument that has a typical coded mask which is similar to that of SWIFT-BAT or SVOM-ECLAIRs.

We compare the sensitivity of these two telescopes for a point source, which is in the center of the FoV, in energy range 0.5-4 keV with equation

\begin{equation}
\label{eq:sensitivity}
F_{\rm min}= n_{\sigma}\cdot\frac{\sqrt{B\cdot A_{\rm B}\cdot\Omega\cdot\varepsilon_{\rm B}}}{\varepsilon_{\rm S} \cdot A_{\rm S}\cdot\sqrt{T\cdot\Delta E}}
\end{equation}
According to the previous discussion, the X-ray background is the predominant component for the MPO telescope and the most X-ray background getting into the detector are from the FoV. Because of the thickness (0.5 mm) of the detector, the energetic cosmic particles cannot cause high background count rate in the low energy range 0.5-4 keV. This conclusion is also true for the coded mask telescope.
In this case, the background estimation in Equation (\ref{eq:sensitivity}) only needs to be X-ray background for the sake of simplicity.

The parameters in Equation (\ref{eq:sensitivity}) for the two telescopes are listed in Table \ref{Tab:Parameters}.
$B$ is the background in unit of cts/s/cm$^2$/sr/keV. We generally calculate sensitivity taking into account the background on the same detector area as the source photons. Therefore, we can take the whole detector area as $A\rm_B$ for coded mask telescope. The solid angle subtended by the coded mask is about 40$\rm^o\times$40$\rm^o$ which is considered as the value of $\Omega$.

For MPO telescope, we collect background information on the focal spot which is marked with a piece of short red line on detector in Figure \ref{Fig:MPOandMask} (left). The solid angle $\Omega_{\rm MPO}$ subtended by focal area is about 0.15$\rm^o\times$0.15$\rm^o$.
We consider the X-rays, whose incidence directions are inside the solid angle $\Omega_{\rm MPO}$, as parallel light. 
According to simulations, when the parallel X-rays are on-axis, the X-rays on the central optical area of 4 cm$\times$4 cm cause about 75\% of the background on the focal area; and the background grows slowly with the optical area when it is larger than 75\%. The central optical area is shaded with green colour in Figure \ref{Fig:MPOandMask} (left).
In other words, most of the background on the focal area is due to the quasi-parallel X-rays which are on the central optical area. 
Therefore, we assume $A\rm_B$ of MPO telescope as 4 cm$\times$4 cm and consider $\Omega$ as 0.15$\rm^o\times$0.15$\rm^o$. 
The corresponding average reflectivity $\varepsilon_{\rm B}$ is approximately 0.1.

$A\rm_{S}$ is the detection area for sources. For the MPO telescope, we consider $A\rm_{S}$ as the effective area of 3 cm$^2$. Thus the X-ray reflectivity $\varepsilon_{\rm S}$ is equal to 1 as $A\rm_{S}$ includes this factor. For the coded mask telescope, $A\rm_{S}$ is the detector area; both $\varepsilon_{\rm S}$ and $\varepsilon_{\rm B}$ are the open ratio of the mask and we assume the value as 0.5. 
$T$ and $\Delta E$ are the observation time and energy range, respectively.

Based on Equation (\ref{eq:sensitivity}) and the parameter values in Table \ref{Tab:Parameters}, we obtain $F\rm_{min}$(coded mask)/$F\rm_{min}$(MPO)$\approx$65. If we take into account the factor 75\% of the background of MPO telescope, we can get a ratio of about 56. In other words, the sensitivity of a MPO telescope is at least one order of magnitude better than that of a coded mask telescope with the same physical dimension. This result is in agreement with our simulations.
Therefore, we can conclude that the result of this comparison for the two different telescopes with the same physical size is reasonable.

The coded mask telescope SWIFT-BAT has a sensitivity of about 5 mCrab \cite{Barthelmy2005} in 10 ks with a 5$\sigma$ for on-axis sources, while the coded mask telescope we discussed above has a sensitivity of 1.3 mCrab. According to our estimation, SWIFT-BAT should have a better sensitivity since it has a much larger mask and detector. However, it is not strict to compare the sensitivity of these two telescopes because SWIFT-BAT works at a harder energy range.

\begin{table}
\begin{center}
\caption{Parameters in Equation (\ref{eq:sensitivity}) for the coded mask telescope and the MPO telescope. These parameters are discussed in the energy range 0.5-4 keV.}
\label{Tab:Parameters}
\small
\begin{tabular}{|c|c|c|c|c|c|}
\hline
 	--		  		 & $A_{\rm B}$(cm$^2$) 	& $\Omega$($\rm$ deg$\times$ deg) 	& $\varepsilon_{\rm B}$  & $\varepsilon_{\rm S}$ 	& $A_{\rm S}$(cm$^2$) \\
\hline
MPO telescope		 &4$\times$4 	&0.15$\times$0.15		&0.1		&1		&3 \\
\hline
coded mask telescope  &14$\times$14		&40$\times$40				&0.5		&0.5		&14$\times$14		\\
\hline
\end{tabular}
\end{center}
\end{table}

\begin{figure}
\centering
\includegraphics[width=1.1\textwidth]{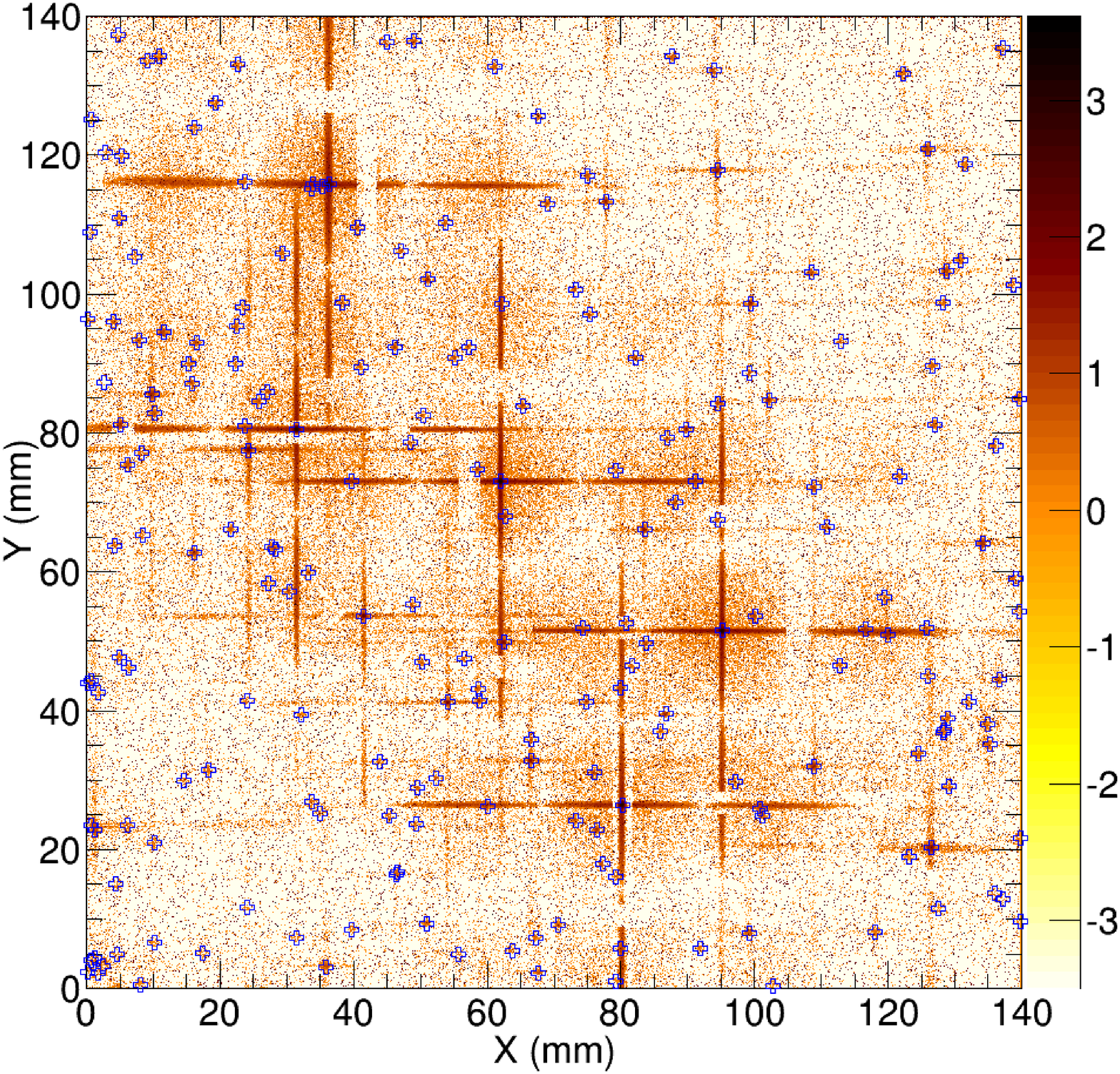} 
\caption{Simulated X-ray image of a sky region (centered at RA=255$\rm^o$ , Dec=-50$\rm^o$) with the telescope (its FoV is 20$\rm^o\times$20$\rm^o$) described in Section \ref{sec:MCmodel} in the energy range 0.5-4 keV and an exposure time 500 ks.
The sources are taken from the RASS Bright Source Catalogue (RASS-BSC) and each blue hollow cross in the figure indicates a source there. Note that background is also included in the data and the color code is on logarithmic scale.
}
\label{Fig:Rosat1}
\end{figure}

\begin{figure}
\centering
\includegraphics[width=0.6\textwidth]{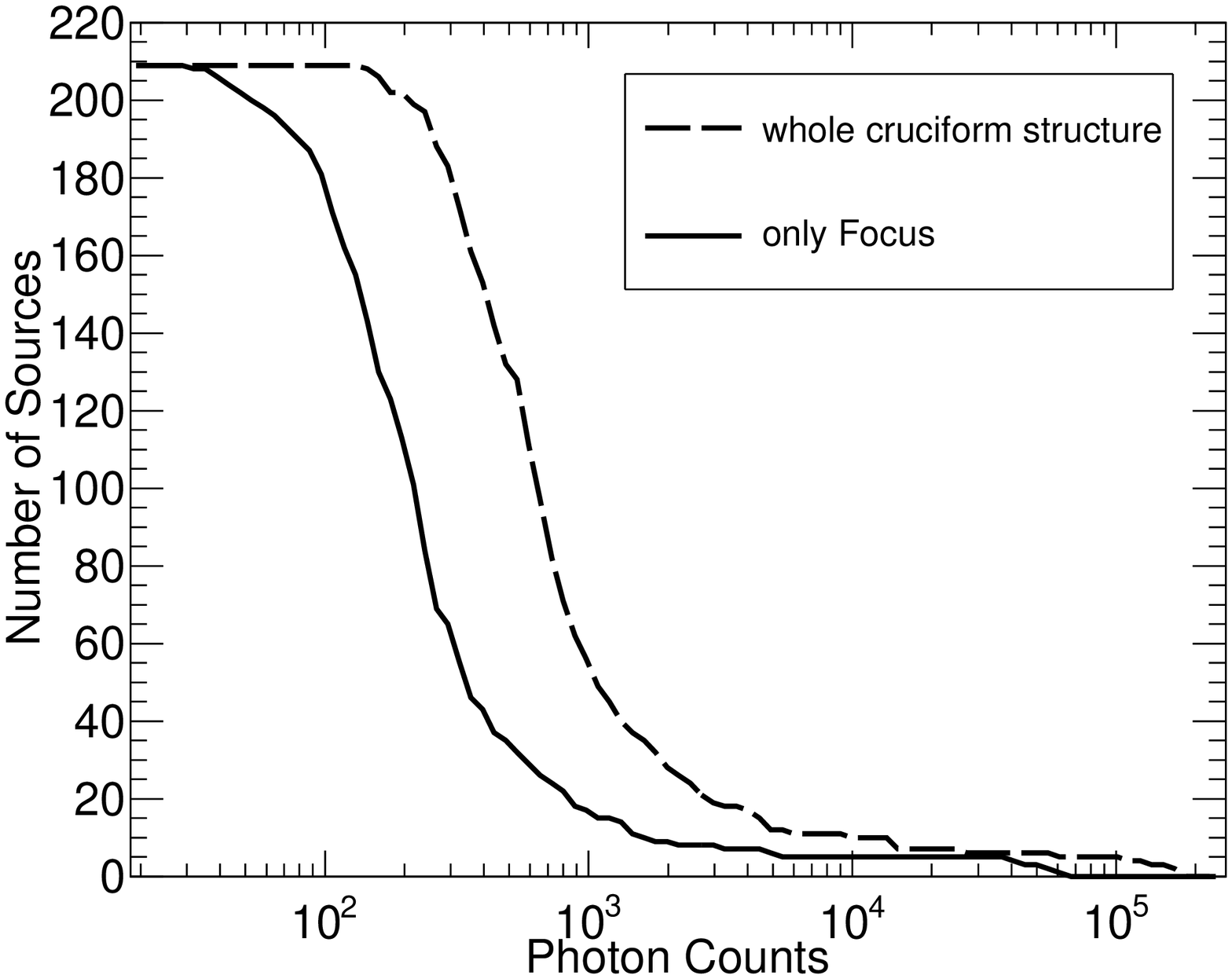}
\caption{The number of sources as a function of the minimum detected photon counts in the RASS-BSC simulation.}
\label{Fig:Rosat2}
\end{figure}

Based on the study of X-ray tracing and the estimation of background,
we simulate the observations of a piece of the X-ray sky with point sources from the ROSAT All-Sky Survey Bright Source Catalogue (RASS-BSC)\footnote{ftp://ftp.xray.mpe.mpg.de/rosat/catalogues/rass-bsc/} \cite{Voges1999}.
We assume a spectral index of 2 and an absorption of 3$\times$10$^{20}$ cm$^{-2}$ for all sources.
Figure \ref{Fig:Rosat1} shows the simulated `observed' X-ray image of the sky region centered at (RA=255$\rm ^o$, Dec=-50$\rm ^o$) within the FoV (20$\rm^o\times$20$\rm^o$), in the energy range 0.5-4 keV, with an exposure time 500 ks.
There are a total of 210 sources inside the FoV and each blue hollow cross in the figure indicates one source.
It can be seen from the brightest sources in Figure \ref{Fig:Rosat1} that there is no noticeable vignetting effect across the FoV. However, the cruciform shape of the PSF may cause confusion for faint sources in the vicinity of bright ones.
As shown in Figure \ref{Fig:Rosat2}, 196 sources can be detected with at least 60 photons in the focal area of their images. It means that about 93\% of RASS-BSC in this part of sky can be detected by the telescope described in Section \ref{sec:MCmodel} in the energy range 0.5-4 keV above 5$\sigma$ detection level in 500 ks. Only one sources at (RA=258.388, Dec= -39.7620) can not be detected because its count rate is zero in RASS-BSC.

%%%%%%%%%%%%%%%%%%%%%%%%%%%%%%%%%%%%%%%%%%%%%%%%%%%%%%%%%%
\section{Summary and Discussion}
\label{sec:sum}

Even though Geant4 is a powerful toolkit which can simulate various interactions of particles and materials, it lacks some necessary physical processes at present. In this paper, we describe the detailed usage of the extensions XRTG4 and FirsovSt of Geant4 and present simulations of a wide-angle X-ray focusing telescope including the X-ray tracing, proton scattering and background simulations.

Our main results for the wide-angle X-ray focusing telescope described in Section \ref{sec:MCmodel} are: (1) we verified that the effective areas obtained with Geant4 and Q software are in a good agreement with an average difference less than 3\%; (2) X-rays are the dominant background sources below 10 keV; (3) the sensitivity of the lobster-eye telescope is expected to be 0.02 mCrab (5 $\sigma$, 10 ks, a power-law spectrum with photon index 2.1 and absorption 3$\times$10$^{20}$ cm$^{-2}$) and is at least one order of magnitude better than that of a coded mask telescope with the same dimensions; and (4) the number of protons passing through optics and reaching the detector with Firsov scattering is about 2.5 times that with MSC.

Based on these investigations with the powerful Geant4, we can build a unified model of grazing incidence X-ray telescopes. We can use the simulations of such telescopes to optimize their design including the shielding geometry, the performance of the detector and the detection
capabilities of light curves and spectra for different kinds of X-ray sources.
The biggest advantage of doing simulations with such a unified model is that we can take into account the whole structure of the telescope including optics, detector, shielding and other structures at the same time. Thus, we can consider all the factors which may affect the scientific performance of the telescope in a comprehensive and accurate analysis.

\section*{Acknowledgements}

This work is supported by the Strategic Priority Research Program on Space Science, the Chinese Academy of Sciences, Grant No. XDA04061100, the programs of National Natural Science Foundation of China under the Grant No.11403055, No.11203043 and No.11427804, China Post-doctoral Science Foundation under the Grant No.2014M561068, and the Young Researcher Grant of National Astronomical Observatories, Chinese Academy of Sciences.

\end{document}